\documentclass[preprint,11pt]{elsarticle}
\usepackage{amsmath,amsthm,amssymb,graphicx}
\usepackage[lined,boxed,commentsnumbered]{algorithm2e}
\usepackage{pgfplots}
\usepackage{comment}
\usepackage{bm}
\usepackage{subcaption}
\usepackage{lscape}
\usepackage{graphicx}
\usepackage{rawfonts}
\input{prepictex}
\input{pictex}
\input{postpictex}
\newcommand{\nam}[1]{{\color{black}{#1}}}

\theoremstyle{definition}

\theoremstyle{remark}
\newcommand{\blue}[1]{{\color{black}{#1}}}

\newcommand{\vs}{\vspace{.5ex}}
\newcommand{\ds}{\displaystyle}
\newcommand{\half}{{\textstyle\frac{1}{2}}}
\newcommand{\gap}{\Delta_{\mathrm{eff}}}

\definecolor{gold}{rgb}{0.85,0.65,0}
\definecolor{dgreen}{rgb}{0, 0.5, 0}

\def\beq{\begin{equation}}
\def\eeq{\end{equation}}

\def\fnote#1{\footnote}

\newtheorem{proposition}{Proposition}

\theoremstyle{definition}

\def\cA{{\cal A}}
\def\cB{{\cal B}}

\newcommand{\CA}{\mathcal{A}}
\newcommand{\CB}{\mathcal{B}}

\renewcommand{\vs}{\vspace{.5ex}}
\renewcommand{\ds}{\displaystyle}
\renewcommand{\half}{{\textstyle\frac{1}{2}}}
\newcommand{\fourth}{{\textstyle\frac{1}{4}}}
\renewcommand{\gap}{\Delta_{\mathrm{eff}}}

\theoremstyle{definition}

\begin{document}

\begin{frontmatter}


\title{Political Districting without Geography}


\author{Gerdus Benad\`e}
\ead{benade@bu.edu}
\address{Boston University}
\author{Nam Ho-Nguyen}
\ead{nam.ho-nguyen@sydney.edu.au}
\address{The University of Sydney}
\author{J. N. Hooker}
\ead{jh38@andrew.cmu.edu}
\address{Carnegie Mellon University}

\begin{abstract}
\blue{Geographical considerations such as contiguity and compactness are necessary elements of political districting in practice.  Yet an analysis of the problem without such constraints yields mathematical insights that can inform real-world model construction.  In particular, it clarifies the sharp contrast between proportionality and competitiveness and how it might be overcome in a properly formulated objective function.  It also reveals serious weaknesses of the much-discussed efficiency gap as a criterion for gerrymandering.}
\end{abstract}




\end{frontmatter}

\section{Introduction}

Optimization models have been devised for political districting for more than half a century (e.g., \cite{HWSWZ65,GN70}).  As observed in \cite{RSS13}, these models have been almost exclusively concerned with the geographical layout of districts, aside from ensuring that districts have roughly equal populations.  
\blue{Contiguity and compactness are seen as particularly important and mandated in 34 and 31 U.S.\ states, respectively \cite{CRS21}.}  
The very term {\em gerrymandering} refers to the salamander-like shape of districts that are contrived to benefit a certain party.  Yet the fundamental problem with gerrymandering is not the shape of the districts, but the unfair representation that results.  The ``packing and cracking'' strategies used in gerrymandering are based on the political demographics of districts, not their geography.  Districts that look reasonable can be highly gerrymandered, while distorted and serpentine districts can provide fair representation.

In recent years, a rise in political polarization has led to concern about the competitiveness of districts as well as gerrymandering \cite{AltMcD15,Cot19,SKJ19,DefDucSol20}.  When individual districts are dominated by a single political party, their representatives may be less inclined to negotiate compromise, possibly resulting in a more partisan legislature.  Yet competitiveness is no more tied to geography than gerrymandering is.  Districts that concentrate a single point of view can be either compact or serpentine.

The nongeographical essence of the fair districting problem suggests that it can be usefully analyzed without the distraction of geographical constraints.  We find that such an analysis, even though it relies solely on elementary algebra, reveals basic properties of the problem that, to our knowledge, have not been observed in the literature.  For example, \blue{it \mbox{reveals} the enormous theoretical potential of gerrymandering to undermine proportional representation, and it} clarifies the conflict between competitiveness and proportional representation.  More importantly, it can lead to optimization models that better incorporate the fundamental goal of \blue{proportional} representation without sacrifice of competitiveness.  

A geography-free analysis also reveals serious weaknesses in the recently much-discussed {\em efficiency gap} criterion for fair districting \cite{SKJ19,SM15,BM17,Cov18}.  The efficiency gap measures the extent to which the political parties differ in how many of their votes are ``wasted.''  We show that minimizing the efficiency gap is consistent with highly  nonproportional representation and extreme noncompetitiveness.  It is therefore unsuitable, we argue, as an objective.

\blue{We recognize that geographical constraints are often a necessary and legitimate component of the districting problem.  Highly contorted districts can raise public suspicions of gerrymandering even if it does not exist.  There may be advantages in preserving the integrity of local political entities such as counties or precincts, and compact districts may facilitate a legislator's task of maintaining contact with constituents.  A practical model must represent these and other complexities of the real-world situation.  Yet our purpose here is not to develop such a model, but to show that removing geographical constraints for purposes of conceptual analysis can reveal fundamental insights into the nature of the districting problem.  These insights can, in turn, inform the design of practical models, particularly the formulation of the objective function.  }

\blue{
An excellent survey of districting models appears in \cite{RSS13}.
More recent models include \cite{VBL20,GS20,OH17,LiuErdLinTsa20}.
Various measures of gerrymandering and competitiveness are discussed in \cite{Nag19}, which provides further references.
In addition to the efficiency gap, many other metrics for evaluating properties of districting plans have been proposed including the geometric target \cite{LS14,GPT21} and partisan symmetry as measured by the mean-median metric or partisan bias \cite{Gro83,KB87,GK07}. Like the efficiency gap, symmetry metrics have flaws \cite{DDD+20}. The (non-)compatibility of these metrics with competitiveness is left for future study. 
}

\section{The Basic Model}


To simplify discussion we assume two political parties, A and B, although our analysis can be readily extended to multiple parties or interest groups.  We let $\alpha$ and $\beta$ be the fraction of the voting population aligned with parties A and B, respectively, where $\alpha+\beta=1$.  The legislature contains $n$ seats, corresponding to $n$ districts.  We suppose that A is the majority party ($\alpha>\beta$), that all districts have the same population, and that every eligible voter votes.
\blue{
The political districting problem in its simplest form is to:
\begin{equation}\label{eq:political-districting}
\begin{aligned}
&\text{decide, for each district $i=1,\ldots,n$,}\\
&\text{the fraction $\alpha_i$ of its voters aligned with party~A.}\\
\end{aligned}\tag{PD}
\end{equation}
The fraction of voters aligned with party~B will be $\beta_i = 1-\alpha_i$. The notation is summarized in Table~\ref{ta:symbols}.

In a practical political districting problem (where geography is considered), the task is to decide the \emph{shape} of each district, or equivalently which geographical regions to assign to each district, which then implicitly determines the fractions $\alpha_i,\beta_i$. These fractions are, in turn, used to assess the appropriateness/fairness of particular districting plans via metrics such as proportionality, competitiveness and efficiency gap. The goal of our paper is to critically examine the merits and pitfalls of these metrics, which we hope will lead to a deeper understanding of the problem, and can inform practical districting. With this in mind, we have chosen to ignore geographic considerations in \eqref{eq:political-districting}, and instead directly focus on determining the fractions $\alpha_i,\beta_i$.
}

\begin{table}[!t]
\centering
\caption{List of Symbols} \label{ta:symbols}
\centering
{\small
\begin{tabular}{ll}
$n$ & number of seats in the legislature \\
$m$ & number of seats won by party A \\
$\alpha$, $\beta$ & fraction of total population that votes for party A, B \\
$\alpha_i$, $\beta_i$ & fraction of population of district $i$ that votes for party A, B \\
$\bar{\alpha}$, $\bar{\beta}$ & average of $\alpha_i$, $\beta_i$ across \mbox{majority-A}, \mbox{majority-B} districts \\
$\CA$, $\CB$ & index set of majority-A, majority-B districts \\
$\rho$ & proportionality ratio for party B: $(1 - m/n)/\beta$ \\
$\Delta$ & voting margin $\alpha-\beta$ in the population as a whole \\
$\delta$ & district-level competitiveness margin \\
$\delta'$ & margin in noncompetitive districts \\
$\gap$ & efficiency gap 
\end{tabular}
}
\end{table}

We first investigate how to design districts so that a given number $m$ of the districts are majority~A.  Let $\cA$ be the index set of \mbox{majority-A} districts, and similarly for $\cB$.  Then since all districts contain the same number of voters, we have
\begin{equation}
\frac{1}{n}\Big(\sum_{i\in \cA} \alpha_i + \sum_{i\in \cB} (1-\beta_i)\Big)= \alpha
\label{eq:00}
\end{equation}
Let $\bar{\alpha}$ be the average fraction of party A adherents in \mbox{majority-A} districts, and similarly for $\bar{\beta}$, so that
\[
\bar{\alpha} = \frac{1}{m}\sum_{i\in \CA} \alpha_i \hspace{5ex}
\bar{\beta} = \frac{1}{n-m}\sum_{i\in \CB} \beta_i
\]
Then (\ref{eq:00}) immediately implies  
\[
\frac{m}{n}\bar{\alpha} + \Big(1 - \frac{m}{n}\Big)(1-\bar{\beta}) = \alpha
\]
From this we have the following.

\begin{proposition} \label{prop:1}
If the districts have equal population, then 
\begin{equation}
\frac{m}{n} = \frac{\alpha+ \bar{\beta} - 1}{\bar{\alpha} + \bar{\beta} -1}
\label{eq:02}
\end{equation}
\end{proposition}

Thus the number of seats allocated to party A is determined by the {\em average} fraction of A voters in \mbox{majority-A} districts and the average fraction of B voters in \mbox{majority-B} districts.  The distribution of A and B voters across their respective majority districts has no effect.

We can also derive bounds on the fractions $\bar{\alpha}$ and $\bar{\beta}$.  We first note that \eqref{eq:02} implies
\begin{equation}
\bar{\alpha} = 
\frac{\begin{array}{@{}c@{}} {\ds \alpha-\Big(1-\frac{m}{n}\Big)(1-\bar{\beta})} \\ \ \vspace{-2ex} \end{array}}
{\begin{array}{@{}c@{}} \ \vspace{-2ex} \\ {\ds \frac{m}{n}} \end{array} } 
\hspace{5ex}
\bar{\beta} = 
\frac{\begin{array}{@{}c@{}} {\ds \beta-\frac{m}{n}(1-\bar{\alpha})} \\ \ \vspace{-2ex} \end{array}}
{\begin{array}{@{}c@{}} \ \vspace{-2ex} \\ {\ds 1 - \frac{m}{n}} \end{array} }
\label{eq:03}
\end{equation}
Due to the fact that $\half<\bar{\beta}\leq 1$, the first equation in (\ref{eq:03}) implies
\begin{equation}
\half + \frac{n}{m} (\alpha-\half) < \; \bar{\alpha} \leq \frac{n}{m}\alpha \label{eq:04} 
\end{equation}
Since the upper bound in \eqref{eq:04} may be greater than 1, we replace it with $\min\{1,(n/m)\alpha\}$.  We substitute into \eqref{eq:04}, so modified, the expression for $\bar{\alpha}$ in \eqref{eq:03} to obtain bounds on $\bar{\beta}$.  This yields
\begin{proposition}
If the districts have equal population, then the fractions $\bar{\alpha}$ and $\bar{\beta}$ have the bounds
\begin{align}
\half + \frac{n}{m} (\alpha-\half) < & \; \bar{\alpha} \leq \min \Big\{1, \;\frac{n}{m}\alpha\Big\} \label{eq:04a} \\
\half < & \; \bar{\beta} \leq \min \Big\{ 1, \; \frac{n}{n-m}\beta \Big\} \label{eq:06}
\end{align}
\end{proposition}
\noindent

As a running example, suppose the electorate consists of 60\% party A supporters ($\alpha=0.6$).  If we wish to allot 7 of 10 legislative seats to party A ($m/n=7/10$), the average fraction $\bar{\alpha}$ of A voters in \mbox{majority-A} districts must be between 64\% and 86\%, from \eqref{eq:04a}.  The resulting average fraction $\bar{\beta}$ of B voters in \mbox{majority-B} districts can be anything between 50\% and 100\%, from \eqref{eq:06}.

\section{Gerrymandering}

The above simple model reveals the \blue{theoretical potential of gerrymandering to defeat proportional representation. Suppose party~B wants to  gerrymander the districts so that it} will win $n-m>n/2$ of the seats and control the legislature, even though it is the minority party.  \blue{It can accomplish this} by cracking and packing.  It {\em cracks} the B vote by letting the \mbox{majority-B} districts have a small average margin $\epsilon$, so that $\bar{\beta}-(1-\bar{\beta})=\epsilon$, or $\bar{\beta}=\half(1+\epsilon)$.  Substituting this into the expression for $\bar{\alpha}$ in (\ref{eq:03}), we have 
\begin{equation}
\bar{\alpha} = \frac{n}{m}\alpha - \half(1-\epsilon)\Big(\frac{n}{m}-1\Big)
\label{eq:20}
\end{equation}
This is the average fraction of A voters that must be {\em packed} into \mbox{majority-A} districts to ensure that $n-m$ districts vote B by an average margin of $\epsilon$.

In the example, suppose party B wishes to win 6 of the 10 seats even though it has only 40\% of the vote.  It need only give the \mbox{majority-B} districts a slight majority of B voters and pack the \mbox{majority-A} districts with slightly more than 75\% A voters on the average.

To find the largest number of B districts the party can engineer (i.e., the largest value of $n-m$), we note that when $\bar{\beta}=\half(1+\epsilon)$, (\ref{eq:02}) implies
\begin{equation}
\frac{m}{n} = \frac{\alpha-\half(1-\epsilon)}{\bar{\alpha}-\half(1-\epsilon)}
\label{eq:07}
\end{equation}
To maximize $n-m$ for a given $n$, we note from (\ref{eq:07}) that the smallest integer $m$ such that $\bar{\alpha}\leq 1$ is
\[
m = \left\lceil n\frac{\alpha-\half(1-\epsilon)}{1-\half(1-\epsilon)} \right\rceil
= n - \left\lfloor \frac{2n\beta}{1+\epsilon} \right\rfloor
\]
Thus the largest value of $n-m$ we can obtain is 
\[
n-m = \left\lfloor \frac{2n\beta}{1+\epsilon} \right\rfloor 
= \left\{
\begin{array}{ll}
\lfloor 2n\beta \rfloor, & \mbox{if $\lfloor 2n\beta\rfloor<2n\beta$} \\
2n\beta - 1, & \mbox{if $\lfloor 2n\beta\rfloor = 2n\beta$}
\end{array}
\right.
\]
where the second equality holds for sufficiently small $\epsilon > 0$.  Also we have $n-m > n/2$ when $\beta>\fourth$.  
Thus 
\begin{proposition}
If the districts have equal population, gerrymandering can yield at least as many as $2\beta n -1$ seats for the minority party when $2n\beta$ is integral and $\lfloor 2\beta n \rfloor$ seats otherwise.  In particular, the minority party can control the legislature if it accounts for more than a quarter of the population.
\end{proposition}

For example, if only 41\% of the population votes for party B, it can gerrymander the districts so as to win 8 of the 10 seats.  In fact, gerrymandering can give party B control of the legislature if it accounts for only 26\% of the population. 

\section{Proportionality and Competitiveness}

Proportionality, or proportional representation, means that the fraction of districts that favor a given party is roughly the fraction of people who belong to that party.  Competitiveness means that the minority party in a district has some chance of winning future elections, which can occur when the fraction of people who belong to it is not too much less than 50\%. 
We will see that competitiveness in all districts is sharply at odds with proportionality.

We define a proportionality ratio $\rho$ to be the ratio of party B's representation in the legislature to its representation in the population, so that $\rho=(1-m/n)/\beta$.  A ratio $\rho=1$ is ideal, while $\rho=0$ means that the minority party wins no seats at all, and $\rho>1$ means it is overrepresented.  Maximizing proportionality corresponds to minimizing $|1-\rho|$.

We measure competitiveness in a district by the margin of that district's majority party over its minority party.  Thus if we require a margin of $\delta$ in every district, we have $\delta=\alpha_i-\beta_i$ in \mbox{majority-A} districts and $\delta=\beta_i-\alpha_i$ in \mbox{majority-B} districts.
This implies
\begin{equation}
\beta = \half - \Big( \frac{m}{n} - \half\Big)\delta, \;\;
\mbox{or} \;\;
\frac{m}{n} = \half + \frac{\half - \beta}{\delta}
\label{eq:023}
\end{equation}
The tradeoff between proportionality and competitiveness is more intuitive when the competitiveness margin $\delta$ is compared to the overall margin $\Delta$ between the parties.  Thus we let $\Delta=\alpha-\beta=1-2\beta$, so that $\beta=\half(1-\Delta)$.  Using this, \eqref{eq:023}, and the definition of $\rho$, we obtain the following, which holds with or without geographical constraints:
\begin{proposition}
If all the districts have the same population, and $\Delta$ is the voting margin in the population as a whole, then a margin of $\delta$ in each district results in a proportionality ratio
\begin{equation}
\rho = \frac{1-\Delta/\delta}{1-\Delta}
\label{eq:010}
\end{equation}
\end{proposition}

This result implies a severe incompatibility between proportionality and general competitiveness.  We first note that $\rho\leq 1$ because $\delta\leq 1$.  Furthermore, we can see as follows that greater competitiveness in all districts (smaller $\delta$) implies much less  proportionality (smaller $\rho$).  Since necessarily $\rho\geq 0$, \eqref{eq:010} implies $\delta\geq\Delta$.  Formula \eqref{eq:010} also reminds us that the minority party wins no seats at all when $\delta=\Delta$.
Now suppose, for example, that the minority party represents $\beta=48\%$ of the voters, so that $\Delta=4\%$.  \blue{If we desire a reasonable proportionality ratio of $\rho=5/6$, which  allows the minority party to win $\rho\beta=40\%$ of the seats, we must tolerate a large margin of $\delta=20\%$ in each district.  If we wish to achieve a more competitive margin of 8\%, then $\rho=52\%$, and the minority party must settle for only $25\%$ of the seats.}  

Thus, even a modest degree of competitiveness excludes any semblance of proportionality \blue{when all districts have the same margin.  We will see in Section~\ref{designing}, however, that by allowing some districts to have larger margins than others, we can arrange for $2(n-m)$ districts to be very competitive.}

\section{Efficiency Gap}\label{sec:efficiency-gap-problems}

The efficiency gap is a much-discussed measure of gerrymandering.  When the gap is small, gerrymandering is presumably less severe, which suggests that a reasonable objective is to minimize the efficiency gap.  However, we will see that there are three problems with minimizing the efficiency gap.  
\begin{itemize}
	\item The efficiency gap is fully determined by the total population of districts won by the majority party.  It is insensitive to any other characteristics of the districting plan.  
	\item Minimizing the efficiency gap is consistent with a substantial lack of proportionality, except when the  two parties have roughly equal support in the population.
	\item Minimizing the efficiency gap is consistent with a complete absence of competitiveness. \blue{Moreover, this can occur \emph{simultaneously} to a lack of proportionality.}
\end{itemize}
It therefore seems desirable to strive for proportionality and competitiveness directly, rather than use the efficiency gap as a measure of fairness.

\subsection{Computing the Efficiency Gap}

The efficiency gap is defined as the absolute difference between the number of votes ``wasted'' by party A and the number wasted by party B, divided by the total number of votes.  The number of votes wasted by party A in a given district is the number of votes cast for A minus the number necessary to win, or if A loses in the district, the total number of votes cast for A in the district; and similarly for B.  

We no longer assume that all districts have equal size, and so the treatment to follow is fully general with respect to the calculation of the efficiency gap.  Let $p_i$ be the population (number of voters) in district $i$.  Let $P$ be the total population, $p_A$ the total population of \mbox{majority-A} districts, and similarly for $p_B$, so that
\[
P=\sum_{i=1}^n p_i  \hspace{5ex} 
p_A = \sum_{i\in\CA} p_i \hspace{5ex}
p_B = \sum_{i\in\CB} p_i \hspace{5ex}
\]
The number of votes wasted by parties A and B, respectively, is given by
\[
\sum_{i\in \CA} (\alpha_i-\half)p_i + \sum_{i\in \CB} \alpha_ip_i \;\;\mbox{and}\;\;
\sum_{i\in \CB} (\beta_i-\half)p_i + \sum_{i\in \CA} \beta_ip_i
\]
The absolute difference is
\[
\Big| \sum_{i\in \CA}(\alpha_i-\beta_i-\half)p_i - \sum_{i\in \CB}(\beta_i - \alpha_i -\half)p_i\Big|
=\Big|\sum_{i=1}^n \alpha_ip_i - \sum_{i=1}^n \beta_ip_i + p_B - \half P \Big|
\]
Dividing by $P$, we obtain the following, which holds with or without geographical constraints or equal district populations:
\begin{proposition}
The efficiency gap is given by
\[
\gap = \Big|\alpha-\beta+\frac{p_B}{P}-\half\Big|
= \Big| \frac{p_B}{P}-2\beta+\half\Big|
\]
Thus for a given $\beta$, the efficiency gap depends only on the fraction of the population that lives in \mbox{majority-B} (or \mbox{majority-A}) districts.  The distribution of A and B voters across individual districts has no influence.   
\end{proposition}

\subsection{Minimizing the Efficiency Gap}

We now consider how to minimize the efficiency gap for a given $\beta$.  The gap is zero when $p_B/P = 2\beta-\half$.  
Since $p_B/P\geq 0$ and $\beta\leq \half$, this minimum can be achieved only when $\fourth\leq \beta \leq \half$.  When $0\leq\beta\leq \fourth$, we must set $p_B/P=0$ to obtain a minimum efficiency gap of $\half-2\beta$.  Thus we have the following, which does not assume equal district populations:
\begin{proposition} \label{prop:minGap}
If there are no geographical constraints, the efficiency gap is minimized when
\begin{equation}
\frac{p_B}{P} = \max\big\{2\beta-\half, 0\big\}
\label{eq:300}
\end{equation}
and the resulting gap is 
\begin{equation}
\gap = \left\{
\begin{array}{ll}
\half-2\beta & \mbox{if $0\leq \beta\leq\fourth$} \vs \\
0 & \mbox{if $\fourth\leq \beta\leq\half$} 
\end{array}
\right.
\label{eq:301}
\end{equation}
\end{proposition}

\noindent
This minimum may not be achievable in the presence of geographical constraints.%
\footnote{{As a real-world example of the effect of geographical constraints, Massachusetts is known to have roughly 30\% Republican voters spread fairly homogeneously throughout the state. This, together with state laws governing redistricting,  makes it impossible to create any districts won by the Republican party \cite{DGH+18} and, correspondingly, leads to a minimum efficiency gap much larger than the bound above.}}

If we assume the districts have equal population, \mbox{$p_B/P = 1-m/n$}.  Thus if $\fourth\leq\beta\leq\half$, we minimize the efficiency gap by choosing $m$ so that $1-m/n$ is as close as possible to $2\beta-\half$.  That is, we set \mbox{$m = \lfloor(\frac{3}{2}-2\beta)n + \half \rfloor$}.  The resulting proportionality ratio is 
\begin{equation}
\rho = \frac{\ds 1-(1/n)\big\lfloor\big({\textstyle\frac{3}{2}}-2\beta\big)n+\half\big\rfloor}
{\beta}
\label{eq:rho}
\end{equation}
If $0\leq\beta\leq\fourth$, we set $m=n$, and the proportionality ratio is $\rho=0$. 

In the example with $\beta=40\%$ and equally sized districts, the efficiency gap is minimized at zero when $m=7$.  We can achieve this gap with any districting plan in which party~B wins $1-m/n=30\%$ of the districts.  The resulting proportionality ratio is $\rho=75\%$, from \eqref{eq:rho}.

\subsection{Proportionality and Competitiveness}

A minimized efficiency gap is consistent with a severe lack of proportionality and competitiveness.   
Supposing again that the districts have equal size, \eqref{eq:rho} implies that the proportionality ratio decreases rapidly with the minority party's share of the population.  For example, if there are 10 districts, the minority party obtains 30\% of the seats when its share is 40\%, but it receives only 10\% of the seats when its share is 30\%, and no seats at all when its share is 25\%.      

A minimized efficiency gap also implies a very large competitiveness margin.  Recall that $p_B/P = 1-m/n$ when the districts have equal size.  If we again suppose that $\delta$ is the competitiveness margin in every district, then we have from (\ref{eq:023}) that 
\[
\delta = \frac{\half-\beta}{\half-p_B/P}
\]
Putting this together with Proposition~\ref{prop:minGap}, we conclude the following.
\begin{proposition}
Suppose that all districts have equal population and competitiveness margin $\delta$.  Then the minimum efficiency gap $\gap$, along with the resulting proportionality ratio $\rho$ and competitiveness margin $\delta$, are as given in Table~\ref{ta:minGap}.  
\end{proposition}
\noindent
A minimum efficiency gap of zero, which occurs whenever $\fourth\leq \beta\leq\half$, results in an extremely large competitiveness margin of 50\%. \nam{Concurrently, the proportionality ratio can range anywhere between $0$ and $1$ when \mbox{$\fourth \leq \beta \leq \half$}. For example, $\beta=30\%$ results in a proportionality ratio of $\frac{1}{3}$ and a competitiveness margin of 50\%. Therefore minimizing efficiency gap can result in \emph{simultaneously} poor proportionality and competitiveness.}

\begin{table}[!t]
    \centering
    \caption{Effect of minimizing the efficiency gap}
    \label{ta:minGap}
    \begin{tabular}{cccc}
        $\beta$ range & $\gap$ & $\rho$ & $\delta$ \\ [0.5ex]
        \hline \\ [-2ex]
        $0\leq \beta\leq \fourth$ & $\half-2\beta$ & 0 & $1-2\beta$ \\ [1ex]
        $\fourth\leq \beta\leq\half$ & 0 & $2-\frac{1}{2\beta}$ & $\half$ \\ [0.75ex]
        \hline
    \end{tabular}
\end{table}


\section{Designing an Objective Function} \label{designing}

We have seen that there is a sharp conflict between proportionality and general competitiveness, with or without geographical constraints.  Minimizing the efficiency gap only exacerbates the problem.  While it has the virtue of a direct concern with gerrymandering rather than geographical features, it can result in substantial disproportionality {\em and} a total lack of competitiveness.   

An escape from this dilemma is to aim for proportionality while achieving competitiveness in {\em some} districts, with possibly wider margins in the remaining districts.  A satisfactory degree of proportionality is consistent with a surprisingly large number of highly competitive districts.  This, in turn, suggests a practical objective for the districting problem.

\begin{figure}[!b]
	\centering

	\beginpicture

\setcoordinatesystem units <2.5pt,5pt> 

\put{

\put{

\setlinear

\put{
	
	\put{District} at 50 8
	\put{Total} at 108 7
	\put{voters} at 108 5
	
\put{$1$} at 5 5 
\put{$2$} at 15 5 
\put{$3$} at 25 5 
\put{$4$} at 35 5 
\put{$5$} at 45 5 
\put{$6$} at 55 5 
\put{$7$} at 65 5 
\put{$8$} at 75 5 
\put{$9$} at 85 5 
\put{$10$} at 95 5

} at 0 20

\put{
	
\put{Party A} at -10 7.5
\put{Party B} at -10 2.5

\plot 0 10 100 10 /
\plot 0  0 100  0 /

\plot 0 0 0 10 /
\plot 10 0 10 10 /
\plot 20 0 20 10 /
\plot 30 0 30 10 /
\plot 40 0 40 10 /
\plot 50 0 50 10 /
\plot 60 0 60 10 /
\plot 70 0 70 10 /
\plot 80 0 80 10 /
\plot 90 0 90 10 /
\plot 100 0 100 10 /

\put{$52$} at 108 7.5
\put{$48$} at 108 2.5

\put{$6$} at  5 7
\put{$6$} at 15 7
\put{$6$} at 25 7
\put{$6$} at 35 7
\put{$6$} at 45 7
\put{$6$} at 55 7
\put{$4$} at 65 8
\put{$4$} at 76 8 
\put{$4$} at 85 8
\put{$4$} at 95 8

\plot 0 4 60 4 /
\plot 60 6 100 6 /

\put{$4$} at  5 2
\put{$4$} at 15 2
\put{$4$} at 25 2
\put{$4$} at 35 2
\put{$4$} at 45 2
\put{$4$} at 55 2
\put{$6$} at 65 3
\put{$6$} at 75 3
\put{$6$} at 85 3
\put{$6$} at 95 3

} at 0 13

\put{
	
\put{Party A} at -10 7.5
\put{Party B} at -10 2.5

\plot 0 10 100 10 /
\plot 0  0 100  0 /

\plot 0 0 0 10 /
\plot 10 0 10 10 /
\plot 20 0 20 10 /
\plot 30 0 30 10 /
\plot 40 0 40 10 /
\plot 50 0 50 10 /
\plot 60 0 60 10 /
\plot 70 0 70 10 /
\plot 80 0 80 10 /
\plot 90 0 90 10 /
\plot 100 0 100 10 /

\put{$5+\epsilon$} at 5 7.5
\put{$5+\epsilon$} at 15 7.5
\put{$5+\epsilon$} at 25 7.5
\put{$5+\epsilon$} at 35 7.5
\put{$6$} at 45 7
\put{$6$} at 55 7
\put{$5-\epsilon$} at 65 7.5
\put{$5-\epsilon$} at 75 7.5
\put{$5-\epsilon$} at 85 7.5
\put{$5-\epsilon$} at 95 7.5

\plot 0 5 40 5 /
\plot 40 4 60 4 /
\plot 60 5 100 5 /

\put{$5-\epsilon$} at 5 2.5
\put{$5-\epsilon$} at 15 2.5
\put{$5-\epsilon$} at 25 2.5
\put{$5-\epsilon$} at 35 2.5
\put{$4$} at 45 2
\put{$4$} at 55 2
\put{$5+\epsilon$} at 65 2.5
\put{$5+\epsilon$} at 75 2.5
\put{$5+\epsilon$} at 85 2.5
\put{$5+\epsilon$} at 95 2.5

\put{$52$} at 108 7.5
\put{$48$} at 108 2.5

} at 0 0

} at 20 0

} at 0 0

\endpicture
	\caption{\blue{Redistribution of voters to obtain 8 highly competitive districts with no change in proportionality.  Relative voting populations by party are indicated.}}   \label{fig:districtingORP}
\end{figure}

\blue{To see how this can occur, recall the previous example in which the minority represents 48\% of the population and wins 40\% of the legislative seats, resulting in a wide 20\% margin in all districts.  This is illustrated in the top half of Fig.~\ref{fig:districtingORP}, in which the minority party wins 4 of 10 districts, resulting in a proportionality ratio of $\rho=5/6$.  We can transfer $10\%-\epsilon$ of the party-A voters in 4 of the majority-A districts to a majority-B district, and replace them with the same number of party-B voters from the majority-B districts.  This results in the voter distribution illustrated in the bottom half of Fig.~\ref{fig:districtingORP}.  Note that party~B still represents 48\% of voters and still wins 4 districts.  Yet 8 districts have become highly competitive, with only 2 districts retaining the 20\% margin.  In general, we can obtain as many as $2(m-n)$ highly competitive districts in this fashion.}

\blue{To verify this algebraically, suppose that} we require a small competitiveness margin of $\epsilon$ in $k$ \mbox{majority-A} districts and $k$ \mbox{majority-B} districts, where \mbox{$k\leq n-m$}, and allow a larger margin of $\delta'$ \blue{in the remaining $n-2k$ districts.  This implies}
\begin{align*}
\blue{\beta} & \blue{\;=  \frac{k}{n}(\half+\epsilon) + \frac{k}{n}(\half-\epsilon)
+ \big(1 - \frac{m}{n} - \frac{k}{n}\big) (\half+\half \delta')
+ \big(\frac{m}{n} - \frac{k}{n}\big) (\half - \half\delta')} \\
& \blue{\;= \half - \big(\frac{m}{n} - \half \big)\delta'}
\end{align*}
Note that $k$ and $\epsilon$ drop out of the formula, and we again obtain \eqref{eq:023} and \eqref{eq:010}, except that $\delta'$ replaces $\delta$.  \blue{Figure~\ref{fig:districtingORP} illustrates why any $k\leq n-m$ yields the same result.  Voters can just as easily be exchanged in fewer than $k=4$ majority-B districts without affecting $\beta$ or $\rho$.  }
We conclude the following:

\begin{proposition}
\blue{If all districts have the same number of voters, then for any $k\leq m-n$ there is a districting plan in which $k$ majority-controlled districts and $k$ minority-controlled districts are competitive with arbitrarily small margins, and in which the proportionality ratio is 
\begin{equation}
\rho = \frac{1-\Delta/\delta'}{1-\Delta}
\label{eq:010a}
\end{equation}
where $\delta'$ is the margin in the remaining $n-2k$ districts.}
\end{proposition}


This is an ideal result that assumes an absence of geographical constraints.  Yet it suggests that a reasonable objective is to maximize proportionality (by minimizing $|1-\rho|$) subject to \blue{a lower bound} on the number of competitive districts controlled by \blue{each party}.  An upper bound can be placed on the acceptable margin $\delta'$ in the remaining, possibly noncompetitive districts.  This allows an optimization model to focus on two primary goals of political districting: \blue{proportional} representation without gerrymandering, and avoidance of excessive polarization.  Neither of these goals is fundamentally geographical in nature.





\end{document}